\documentclass[aps,prl,twocolumn,showpacs,superscriptaddress,groupedaddress]{revtex4}  
\usepackage{graphicx}  
\usepackage{dcolumn}   
\usepackage{bm}        
\usepackage{amssymb}   
\usepackage{wasysym}
\begin{document}


\title{ Thermally excited multi-band conduction in LaAlO$_{3}$/SrTiO$_{3}$ heterostructures exhibiting magnetic scattering  }
\author{V. K. Guduru}
\email{guduru@science.ru.nl}
\affiliation{High Field Magnet Laboratory and Institute for Molecules and Materials, Radboud University Nijmegen, 6525 ED Nijmegen, The Netherlands.}
\author{A. McCollam}
\affiliation{High Field Magnet Laboratory and Institute for Molecules and Materials, Radboud University Nijmegen, 6525 ED Nijmegen, The Netherlands.}
\author{A. Jost}
\affiliation{High Field Magnet Laboratory and Institute for Molecules and Materials, Radboud University Nijmegen, 6525 ED Nijmegen, The Netherlands.}
\author{S. Wenderich}
\affiliation{Faculty of Science and Technology and MESA+ Institute for Nanotechnology, University of Twente, 7500 AE Enschede, The Netherlands.}
\author {H. Hilgenkamp}
\affiliation{Faculty of Science and Technology and MESA+ Institute for Nanotechnology, University of Twente, 7500 AE Enschede, The Netherlands.}
\author{J. C. Maan}
\affiliation{High Field Magnet Laboratory and Institute for Molecules and Materials, Radboud University Nijmegen, 6525 ED Nijmegen, The Netherlands.}
\author {A. Brinkman}
\affiliation{Faculty of Science and Technology and MESA+ Institute for Nanotechnology, University of Twente, 7500 AE Enschede, The Netherlands.}
\author{U. Zeitler}
\email{u.zeitler@science.ru.nl}
\affiliation{High Field Magnet Laboratory and Institute for Molecules and Materials, Radboud University Nijmegen, 6525 ED Nijmegen, The Netherlands.}

\date{\today}

\begin{abstract}
Magnetotransport measurements of charge carriers at the interface of a LaAlO$_{3} $/SrTiO$_{3}$ heterostructure with 26 unit cells of LaAlO$_{3} $ show Hall resistance and magnetoresistance which at low and high temperatures is described by a single channel of electron-like charge carriers. At intermediate temperatures, we observe non-linear Hall resistance and positive magnetoresistance, establishing the presence of at least two electron-like channels with significantly different mobilities and carrier concentrations. These channels are separated by 6 meV in energy and their temperature dependent occupation and mobilities are responsible for the observed transport properties of the interface. We observe that one of the channels has a mobility that decreases with decreasing temperature, consistent with magnetic scattering in this channel.
\end{abstract}

\pacs{73.20.-r,73.50.Fq,73.50.Pz}
\keywords{non-linear Hall resistance, positive magneto-resistance, two-band model, oxide heterostructures}
\maketitle

The study of fundamental physical properties and potential applications of complex oxide heterostructures is a rapidly developing field of research \cite{Hwang_NM12,Schl_Sci10,Elis_PRL13}. Interest is largely focussed on the conducting interface between two band-insulating perovskite oxides SrTiO$_{3}$ (STO) and LaAlO$_{3} $ (LAO) \cite{Hwang_Nat04}, which exhibits properties such as superconductivity \cite{Rey_Sci07}, magnetism \cite{Brink_NM07,Ari_NC11,Diki_PRL11,Li_NP11,Bert_NP11,NM13Lee}, and tunable switching of high mobility interface conductivity \cite{Scie06Thie,Nat08Cavig,APL13Gudu}. Although several mechanisms \cite{NM06Nak,PRL09Sing,PRL07Her,PRL07Will} are proposed to be responsible for the interface conductivity, the exact origin and nature of the charge carriers at the interface is still under debate. The major difficulty in achieving consensus about the intrinsic electronic nature of the interfaces is due to the fact that their properties strongly depend on external factors such as the growth conditions of the LAO layer \cite{Brink_NM07}, LAO layer thickness \cite{APL09Bell,PRB10Wong}, and the configuration of the heterostructures \cite{AFM13Huij,NM06Huij}. In order to realize the full potential of LAO/STO heterostructures in technological applications \cite{Schl_Sci10,Elis_PRL13}, the fundamental physical nature of the interface conductivity has to be understood thoroughly.

In this work we study the interface electronic structure of one specific type of LAO/STO heterostructure, which exhibits magnetic signatures \cite{Brink_NM07}. We have performed transport experiments with the magnetic field oriented perpendicular and parallel to the interface, and measured the Hall resistance and sheet resistance in a wide temperature range and at high magnetic fields compared to previous magnetotransport reports \cite{Brink_NM07,Ari_NC11,PRB11Wang,PRL09Bell,PRL10Pent,PRB12Her}. We observed a strong, temperature-dependent, non-linear Hall resistance accompanied by a large positive magnetoresistance (MR) at intermediate temperatures, and a negative MR at low-temperatures \cite{JKPS13Gud}. We quantitatively analyse our data using a simple two-carrier model, considering two electron-like conduction channels with different densities and mobilities. Our interpretation is in line with recent observations that two-channel (electron-like) conduction can be realized in similar LAO/STO heterostructures by means of UV-illumination on the surface of sample at low-temperature \cite{APL13Gudu}. Furthermore, we show that the negative MR \cite{Brink_NM07} at low temperatures is not a result of two-band conduction alone. Rather, we observe that one of the channels has a mobility that decreases with decreasing temperature, consistent with the previously suggested \cite{Brink_NM07} magnetic scenario in this channel.


The sample used for our measurements was grown by pulsed laser deposition using a single-crystalline LaAlO$_{3} $ target. The 10 nm (26 unit cells) LAO film was deposited on a 5 mm $\times$ 5 mm TiO$_{2}$-terminated single crystal STO [001] substrate \cite{APL98Kost}, at a substrate temperature of 850$^\circ$C and an oxygen pressure of $ 2 \times 10^{-3} $ mbar. The growth of the LAO film was monitored using in-situ reflection high-energy electron diffraction, indicating that layer-by-layer growth of individual LAO unit cells (uc) is preserved up to 26 uc. After the growth, the sample was cooled to room temperature with the oxygen pressure remaining at $ 2 \times 10^{-3} $ mbar.

\begin{figure}[ht!]
\includegraphics[width=0.7\columnwidth]{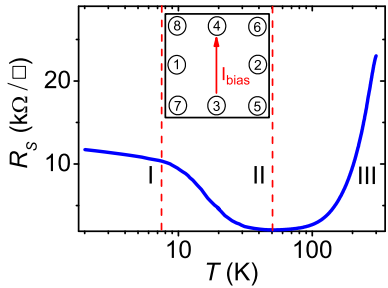}
\caption{ (color online) The sheet resistance $R_{s}$ of the sample as a function of temperature (on a log-scale) between 2 K and 300 K, separated by dashed lines in three regimes. The inset shows the schematic top-view of the sample, with the contact configuration indicated as circles and the bias current I$_{bias}$  with an arrow. }
\label{figure1}
\end{figure}

The sample was mounted on a ceramic chip carrier and electrical contacts were made by ultrasonically bonding aluminium wires. The sample resistance was measured with the contacts as schematically depicted in the inset of Fig. 1: we denote the resistance as $R_{ij,kl}$, where current is passed through the contacts i and j, and the voltage drop is measured between k and l. For a homogeneous sample, the sheet resistance $R_{s}$ is calculated from $R_{34,56}$ using the relation $R_{s} = 1.5 \times R_{34,56}$. We have determined the numeric ratio between $R_{s}$ and $R_{34,56}$ by van der Pauw measurements on different contact configurations where we find similar values for $R_{s}$.

The magnetotransport measurements were performed in a temperature controlled $^{4}$He flow cryostat and in a pumped $^{3}$He system at magnetic fields up to 30 T. The sheet resistance and Hall resistance ($R_{xy} = R_{34,12}$) were measured for both positive and negative magnetic field directions, using a standard, low-frequency lock-in technique with an excitation current of 1 $\mu$A. In order to exclude admixtures of $R_{s}$ in $R_{xy}$ or vice versa we always show the anti-symmetrised Hall resistance data and symmetrized sheet resistance data in the remainder of this paper. The results we report here have been reproduced on several similar samples.


The measured sheet resistance $R_{s}$ is shown as a function of temperature between 2 K and 300 K in Fig. 1. Three distinct regimes can be observed in $R_{s}(T)$ : In region I, from 2 K to 7.5 K, the sheet resistance decreases logarithmically with increase in temperature up to 7.5 K, which is attributed to the Kondo effect originating from the scattering of mobile carriers off localized magnetic moments \cite{Brink_NM07} (our measurements do not extend to low enough temperature to see saturation of this effect); in region II, from 7.5 K to 50 K, further increase of temperature leads to a sharp decrease of resistance with a minimum value around 50 K; in region III, above 50 K, the sample resistance increases monotonically with temperature, which can be attributed to electron-phonon scattering.

\begin{figure}[ht!]
\includegraphics[width=0.8\columnwidth]{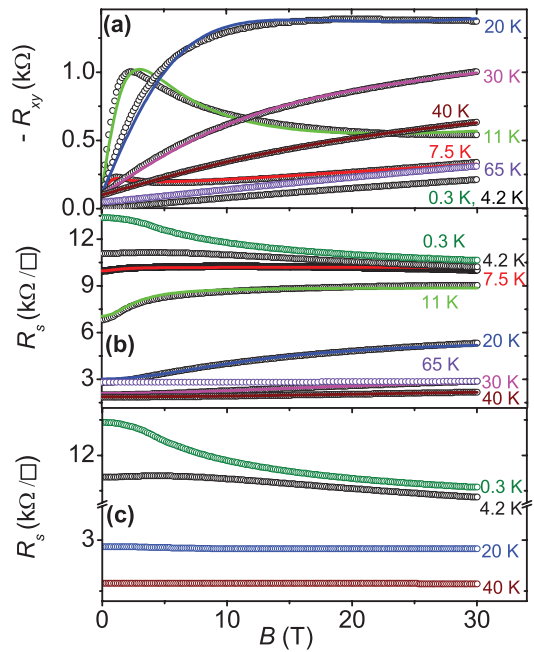}
\caption{ (color online)  (a) Hall resistance $R_{xy}$ and (b) sheet resistance ${R_{s}}$ with the magnetic field oriented perpendicular to the LAO/STO interface for temperatures between 0.3 K and 65 K, shown as open circles. Solid lines in (a) and (b) are the results obtained after simultaneously fitting the $R_{xy}$ and ${R_{s}}$ data using a simple two-carrier model. The $R_{xy}$ curves are offset vertically for clarity. (c) Sheet resistance ${R_{s}}$  for the applied magnetic field oriented parallel to the LAO/STO interface for temperatures between 0.3 K and 40 K, shown as open circles. ${R_{s}}$ axis is cut between 4 and 10 k$\Omega$.}
\label{figure2}
\end{figure}

Fig. 2(a) and 2(b) show the Hall resistance and sheet resistance, respectively, in the magnetic field applied perpendicular to the LAO/STO interface, for temperatures between 0.3 K (region I) and 65 K (region III). At low-temperatures, below 7.5 K, the Hall resistance is linear and independent of temperature, and the sheet resistance decreases as a function of magnetic field, resulting in a strong negative MR. At intermediate temperatures, from 7.5 K to 50 K, the Hall resistance is strikingly non-linear and the sheet resistance increases strongly as a function of magnetic field, resulting in a large positive MR. At high-temperatures, above 50 K, the Hall resistance is linear and the sheet resistance increases slightly as a function of magnetic field, resulting in a negligible positive MR. We also observed a similar non-linear Hall resistance accompanied by a large positive MR, in several other LAO/STO samples with different thickness (5, 10, 15, and 20 unit cells) of LAO film at low-temperature (4.2 K).

Observations of a temperature dependent MR (positive and/or negative) and (non-linear) Hall resistance have been well documented in many different materials. For example, in doped conventional semiconductors such as n-type Ge, GaAs, InSb and Mn doped GaAs \cite{PR64Wood,PR67Kata,PR68Hal,PRB98Mats}, in noble metals doped with transition elements, such as Au doped with Co \cite{Phy59Ger}, and in (magnetic) semiconductor heterostructures (AlGaAs/GaAs, InMnAs/GaAlSb) \cite{PRB88Hout,PRB99Oiwa}, as well as in perovskite oxide heterostructures (LaTiO$_{3} $/SrTiO$_{3}$, LaVO$_{3} $/SrTiO$_{3}$, LaNiO$_{3} $/SrTiO$_{3}$) \cite{PRB10Kim,PRL07Hott,PRL11Sche}. A temperature dependent crossover from negative to positive MR has also been observed in doped semiconductors (In doped CdS) \cite{PRB70Khos} and in magnetic semiconductor heterostructures (InMnAs/GaAlSb) \cite{PRB98Mats,PRB99Oiwa}. The general consensus is that positive MR and non-linear Hall resistance arise due to the contribution to transport of two parallel channels of charge carriers with different mobility. The change in resistance is attributed to the change of electron distribution and mobility in these two channels, caused by the temperature or magnetic field. Negative MR is largely attributed to scattering of conduction charge carriers with localized magnetic moments: the external magnetic field reduces this scattering and results in a decrease of resistance \cite{PRB70Khos}.

Fig. 2(c) shows the sheet resistance data for the applied magnetic field orientation parallel to the LAO/STO interface, for temperatures between 0.3 K (region I) and 40 K (region II). At low-temperatures, the sheet resistance decreases as a function of magnetic field, resulting in a strong negative MR. The MR below 4.2 K is independent of the field orientation, at odds with an interpretation in terms of weak-localization, and consistent with electron spin scattering off localized magnetic moments \cite{Brink_NM07}. The decrease in magnitude of the negative MR with increased temperature can be attributed to the de-localization of magnetic moments by thermal excitation at higher temperatures \cite{PRB11Wang,PRL11Lee}. At intermediate temperatures, the sheet resistance is almost field independent.

The linear Hall resistance below 7.5 K (region I) and above 50 K (region III) can be described using the conventional single-carrier model. The carrier concentration  $(n_{s} = B/R_{xy}e)$ and mobility  $(\mu = 1/ R_{s}(0) e n_{s})$ are extracted from the slope of the linear Hall resistance data and zero field sheet resistance $ R_{s}(0)$ with ${e}$ as the electronic charge. The values obtained for $n_{s}$ and $\mu$ by the single-carrier model are shown in Fig. 3(a) and Fig. 3(b).

In contrast, the non-linear Hall resistance and positive MR between 7.5 K and 50 K (region II) cannot be explained within a single-carrier model, but rather suggest a multi-channel system. A similar non-linear Hall resistance and positive MR were observed previously in LaTiO$_{3}$/SrTiO$_{3}$ \cite{PRB10Kim}, and explained in terms of two-channel conduction from electronic bands with different mobilities, $\mu_{1,2}$ and carrier densities, $n_{1,2}$. We use similar two-electron-band expressions \cite{BCP76Ash},

\begin{equation}
\ {R_{xy}(B)} = \frac{B}{e}  \frac {({n_{1}\mu_{1}^{2}}+{n_{2}\mu_{2}^{2}}) + (\mu_{1}\mu_{2}{B})^2(n_{1}+n_{2})} {({n_{1}\mu_{1}}+{n_{2}\mu_{2}})^2 + {(\mu_{1}\mu_{2}{B})^2(n_{1}+n_{2})^2}} {,}
\label{eq:one}
\end{equation}

\begin{equation}
{R_{s}(B)} = R_{s}(0)\left[1 + \frac {(n_{1}\mu_{1} n_{2}\mu_{2}(\mu_{1}-\mu_{2})^{2}{B}^{2})} {(n_{1}\mu_{1} + n_{2}\mu_{2})^{2}+( (n_{1}+n_{2})\mu_{1}\mu_{2}{B})^{2} }\right] {,}
\end{equation}
\newline
to model our non-linear Hall resistance ${R_{xy}(B)}$ and sheet resistance ${R_{s}(B)}$ data as a function of magnetic field.

\begin{figure}[ht!]
\includegraphics[width=0.7\columnwidth]{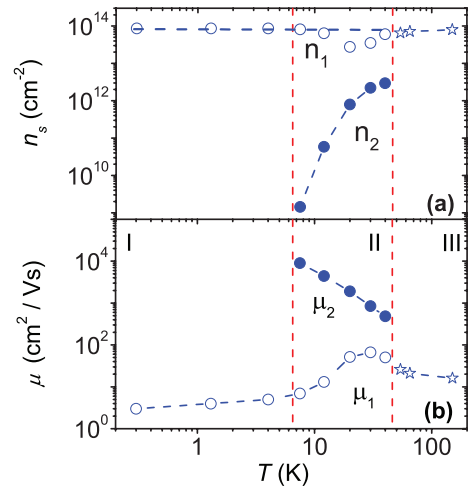}
\caption{(color online) The temperature dependence of sheet carrier density $n_{s}$ and mobility $\mu$ obtained from the analysis of experimental data in Fig. 2(a) and 2(b). (a) The sheet carrier density $n_{s}$ for region I $\&$ II are shown as open circles ($n_{1}$), filled circles ($n_{2}$) and for region III are shown as stars. (b) The carrier mobility $\mu$ for region I $\&$ II is shown as open circles ($\mu_{1}$), filled circles ($\mu_{2}$) and for region III is shown as stars. The connecting dashed lines between the data points are guides to the eye. }
\label{figure3}
\end{figure}

In this expression, we take $n_{1}$, $\mu_{1}$ to be the carrier density and mobility of the existing electron band at low-temperatures and $n_{2}$, $\mu_{2}$ are the carrier density and mobility of the thermally activated second electron band. For low magnetic fields, where $\mu_{1}{B}$, $\mu_{2}{B}$ $\ll ~ 1$, ${R_{xy}(B)}$ shows $B$-linear behaviour, and ${R_{s}(B)}$ shows quadratic dependency on $B$. For high magnetic fields, where $\mu_{1}{B}$, $\mu_{2}{B}$ $\gg ~1$, ${R_{xy}(B)}$ shows $B$-linear behaviour, and ${R_{s}(B)}$ saturates. In field region where $\mu_{2}{B}$ $\simeq 1$ or $\mu_{1}{B}$ $\simeq 1$, ${R_{xy}(B)}$ shows strongly non-linear behaviour.

Fits of the Hall resistance and sheet resistance to this two-band model are shown in Fig. 2(a) and 2(b) as solid lines. The non-linear Hall resistance ${R_{xy}(B)}$ and sheet resistance ${R_{s}(B)}$ at each temperature were fitted simultaneously, using $n_{1}$, $n_{2}$ and $\mu_{1}$, $\mu_{2}$ as fit parameters. The fits are in good agreement with the experimental data and the model is able to nicely reproduce the details of the magnetotransport (${R_{xy}(B)}$ and ${R_{s}(B)}$) in the temperature region between 7.5 K to 50 K, where the data cannot be described within a single carrier model.

The results of our analysis of the magnetotransport data in the three temperature regions are shown in Fig. 3, with the obtained values for the fit parameters $n_{1}$, $n_{2}$ in Fig. 3(a) and $\mu_{1}$, $\mu_{2}$ in Fig. 3(b).

\begin{figure}[ht!]
\includegraphics[width=\columnwidth]{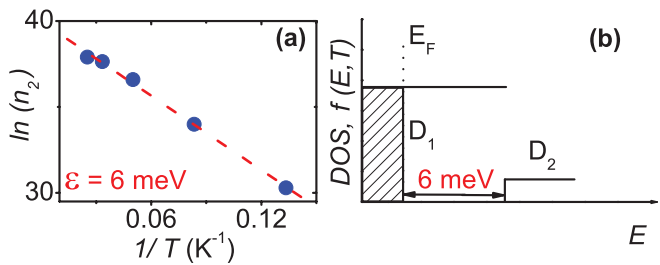}
\caption{ (color online) (a) Arrhenius plot of the thermally excited carrier density $n_{2}$. The slope of the dashed line gives $\varepsilon$ = 6~meV. (b) Schematic representation of the 2D level structure with constant density of states for the two parallel electron-like conduction channels at the interface. At $T$ = 0 K, the level $D_1$ is occupied up to the Fermi energy $E_F$ (dotted line); at $T$ = 60 K, broadening (6 meV) of the Fermi function results in the occupation of level $D_2$. }
\label{figure4}
\end{figure}

For temperatures below 7.5 K (region I), transport is dominated by a low mobility electron band with a temperature independent linear Hall resistance yielding a constant, high, sheet density of $n_{1}$ = $ 8.7 \times 10^{13} $ cm$^{-2}$. The mobility of electrons in this band increases from $\mu_{1}$ = 3 cm$^{2}$/Vs to 5 cm$^{2}$/Vs with an increase in temperature from 0.3 K to 4.2 K. This temperature dependent mobility can be attributed to magnetic scattering which also explains the observed negative MR \cite{Brink_NM07}.

For temperatures between 7.5 K and 50 K (region II), electrons are thermally activated into a second electron band ($n_{2}$, $\mu_{2}$) in addition to the existing low mobility electron band ($n_{1}$, $\mu_{1}$). The carrier density $n_{1}$ in the low mobility electron band stays almost temperature independent while the mobility $\mu_{1}$ increases by about an order of magnitude from 7 cm$^{2}$/Vs to 50 cm$^{2}$/Vs. The carrier density $n_{2}$ in the second electron band shows a strong temperature dependence and increases by a few orders of magnitude from $ 1.4 \times 10^{9} $ cm$^{-2}$ to $ 2.9 \times 10^{12} $ cm$^{-2}$, the mobility $\mu_{2}$ decreases by an order of magnitude from $ 8 \times 10^{3} $ cm$^{2}$/Vs to $ 4 \times 10^{2} $ cm$^{2}$/Vs. The decrease in mobility of $\mu_{2}$ as the temperature increases is likely to be the result of increased electron-phonon scattering. At low temperature $\mu_{1}$ is dominated by magnetic scattering, which is expected to decrease with increasing temperature leading to the observed increase in $\mu_{1}$. As the temperature is further increased, electron-phonon scattering also begins to strongly affect the lower energy conduction channel, so that $\mu_{1}$ reaches a maximum (at $\sim$ 30 K) and then decreases at higher temperatures.

This thermal activation of electrons into a second energetically higher channel is visualized in Fig. 4a. The activation energy of carriers $\varepsilon$ is related to the concentration in the second electron channel by the Arrhenius relation, $n_{2} \propto exp{(-\varepsilon/k_{B}T)}$, where $k_{B}$ is the Boltzmann constant and $T$ is temperature. The linear slope of $\ln (n_{2})$ versus 1/\textit{T} gives $\varepsilon$ = 6~meV. For temperatures above 50 K (region III), the mobilities of the two electron channels become comparable and the two channels no longer give distinguishable contributions to the transport (see expressions 1 and 2). The \textit{total} carrier concentration and mobility are then obtained from a single band calculation. The total carrier concentration remains temperature independent and the average mobility further decreases with increasing temperature.

We explain our results by tentatively considering two simplified interface electronic states with a constant 2D density of states $D_1$ and $D_2$ as schematically depicted in Fig. 4b. The oxygen vacancies in STO (each oxygen vacancy donates 2 free electrons) and/or the electronic reconstruction (half an electron transfer from LAO to STO) give a reservoir of charge carriers at the interface, which includes both localized and mobile carriers. At finite (low) temperature (region I), the charge carriers are present in a lower energy level ($D_1$) which has a low mobility ($\mu_{1}$) and high electron density ($n_{1}$). These carriers are responsible for the observed Kondo type behaviour (i.e., logarithmic increase of sheet resistance accompanied by the decrease of carrier mobility for decreasing temperature down to 0.3 K), negative MR and linear Hall resistance. On increasing the temperature through region II, broadening of the Fermi function results in the thermal excitation of charge carriers into a higher energy level ($D_2$) separated from the lower level by 6 meV, which has a high mobility ($\mu_{2}$) and low electron density ($n_{2}$). Both the low and high mobility bands contribute to the transport in region II and are responsible for the observed large, positive MR and non-linear Hall resistance. When the temperature is further increased above 50 K (region III), the mobilities of these two conduction channels become similar. As a result we can not distinguish the two-band conduction any longer, which eventually gives rise to a linear Hall resistance and a negligible MR.

Our results and analysis strongly suggest that at least two channels of electron-like carriers at the interface are responsible for the observed transport behaviour in these LAO/STO heterostructures exhibiting magnetic scattering, but they do not allow us to determine the exact physical origin of these interface conduction channels. However, from similarities between our results and available information in the literature, we can propose a few possibilities for the origin of the observed electron bands. It has been predicted theoretically \cite{PRL07Pent} and shown experimentally \cite {NM13Lee} that the magnetism in LAO/STO originates from the $t_{2g}$ band of Ti-$3d$ orbitals, specifically from the energy level formed from $d_{xy}$ orbitals with $Ti^{3+}$ character. This suggests that in our sample the observed energy level ($D_1$) associated with the low mobility carriers ($n_{1}$, $\mu_{1}$) and responsible for magnetic effects (Kondo effect, negative MR), could originate from these $d_{xy}$ orbitals. The observed energy level ($D_2$) associated with the high mobility carriers ($n_{2}$, $\mu_{2}$) is separated by 6 meV from the low mobility level in our sample; this activation energy is strikingly similar to values found in earlier observations of carrier activation in SrTiO$_{3}$/LaAlO$_{3}$/SrTiO$_{3}$ heterostructures \cite{NM06Huij,AFM13Huij}, and comparable to La doped SrTiO$_{3}$ \cite{PRB01Oku}. There is evidence from previous work that the higher energy conduction channel we observe is likely to have primarily $d_{xz/yz}$ character \cite{PRL09Sall,PRL13Bern,PRB12Khal}. This seems to be a reasonable assumption for our sample, based on the energy separation of the conduction channels and the carrier density involved \cite{PRB12Khal,Arix13Heer}.

We have also performed similar measurements on LAO/STO samples with thinner LAO layers (20, 15, 10, 5 uc of LAO) where we observe 2-channel conduction of a high mobility and a low mobility channel, which in contrast to the 26 LAO system, is present down to the lowest temperatures. We interpret this observation by a different energetic alignment of the two channels. This coexistence at low temperature precludes a clear distinction of their individual contribution to magnetotransport as compared to the 26 LAO system where the two channels are well separated.

In summary, we have performed magnetotransport experiments on a magnetic LaAlO$_{3}$/SrTiO$_{3}$ interface, with a 10 nm (26 unit cells) LaAlO$_{3}$ film, in magnetic fields up to 30 T. Our experimental results show that the low-temperature regime (\textit{T} $\leq$ 4.2 K) is dominated by a single charge carrier type with a low mobility, yielding a linear Hall resistance and negative MR. Increasing the temperature above 4.2 K leads to a significant decrease of the resistance, a strong positive MR appears, and the Hall resistance becomes distinctly non-linear. Our observations are quantitatively explained by thermal excitation of an additional high mobility electron channel situated 6 meV above the low mobility channel.

\begin{acknowledgments}
This work has been performed at the HFML-RU/FOM member of the European Magnetic Field Laboratory (EMFL) and is part of InterPhase programme supported by the Foundation for Fundamental Research on Matter (FOM) with financial support from the Netherlands Organisation for Scientific Research (NWO).
\end{acknowledgments}

\end{document}